\documentclass[aps,prc,twocolumn,superscriptaddress]{revtex4}

\usepackage{graphics}
\usepackage{longtable}
\usepackage{dcolumn}
\usepackage{amsmath}

\newcommand{\isotope}[3]{{}_{#3}^{#2}\text{{#1}}}
\newcommand{\sm}{\isotope{Sm}{152}{}}
\newcommand{\Pm}{\isotope{Pm}{152}{}}
\newcommand{\gd}{\isotope{Gd}{154}{}}
\newcommand{\eu}{\isotope{Eu}{152}{}}
\newcommand{\pb}{\isotope{Pb}{208}{}}

\newcommand{\git}{School of Physics, Georgia Institute of Technology, Atlanta,
Georgia 30332-0430, USA}

\newcommand{\rochester}{Nuclear Structure Research Laboratory, Department of
Physics, University of Rochester, Rochester, New York, 14627, USA}
\newcommand{\guelph}{Department of Physics, University of Guelph, Guelph,
Ontario N0B 1S0, Canada}
\newcommand{\triumf}{TRIUMF, 4004 Wesbrook Mall, Vancouver, British Columbia
V6T 2A3, Canada}
\newcommand{\ukyp}{Department of Physics and Astronomy, University of Kentucky,
Lexington, Kentucky  40506-0055, USA}
\newcommand{\ukyc}{Department of Chemistry, University of Kentucky,
Lexington, Kentucky  40506-0055, USA}
\newcommand{\ncsu}{Department of Physics, North Carolina State University, Raleigh, North Carolina, 27695-8202, USA}
\newcommand{\llnl}{Lawrence Livermore National Laboratory, Livermore, California 94551, USA}
\newcommand{\richmond}{Department of Physics, University of Richmond, Richmond, Virginia 23173, USA}

\begin{document}


\title{Search for Intrinsic Collective Excitations in $^{152}$Sm}


\author{W.~D.~Kulp}
\author{J.~L.~Wood}
\affiliation{\git}
\author{P.~E.~Garrett}
\affiliation{\guelph}
\affiliation{\triumf}
\author{C.~Y.~Wu}
\altaffiliation[Present address:  ]{\llnl}
\affiliation{\rochester}
\author{D.~Cline}
\affiliation{\rochester}
\author{J.~M.~Allmond}
\altaffiliation[Present address:  ]{\richmond}
\affiliation{\git}
\author{D.~Bandyopadhyay}
\altaffiliation[Present address:  ]{\triumf}
\affiliation{\ukyp}
\author{D.~Dashdorj}
\altaffiliation[Present address:  ]{\llnl}
\affiliation{\ncsu}
\author{S.~N.~Choudry}
\affiliation{\ukyp}
\author{A.~B.~Hayes}
\affiliation{\rochester}
\author{H.~Hua}
\affiliation{\rochester}
\author{M.~G.~Mynk}
\affiliation{\ukyc}
\author{M.~T.~McEllistrem}
\affiliation{\ukyp}
\author{C.~J.~McKay}
\affiliation{\ukyp}
\author{J.~N.~Orce}
\affiliation{\ukyp}
\author{R.~Teng}
\affiliation{\rochester}
\author{S.~W.~Yates}
\affiliation{\ukyp}
\affiliation{\ukyc}



\begin{abstract}
  The 685 keV excitation energy of the first excited $0^+$ state in $\sm$ makes it an attractive candidate to explore expected two-phonon excitations at low energy.
  Multiple-step Coulomb excitation and inelastic neutron scattering studies of $\sm$ are used to probe the $E2$ collectivity of excited $0^+$ states in this ``soft''  nucleus and the results are compared with model predictions.
  No candidates for two-phonon $K^\pi=0^+$quadrupole vibrational states are found.
  A $2^+$, $K=2$ state with strong $E2$ decay to the first excited $K^\pi = 0^+$ band and a probable $3^+$ band member are established. \\
  Accepted for publication as a Rapid Communication in Physical Review C, http://prc.aps.org/.  Copyright (2008) by the American Physical Society.
\end{abstract}


\maketitle


  Low-energy collective structure in nuclei is a fundamental manifestation of simple behavior in finite many-body quantum systems.  
  Nuclear collectivity is divided into two basic types:  rotational (exhibited in the band structures of deformed nuclei) and vibrational (suggested to be dominant in spherical nuclei)\cite{Bohr1975b}.
  Deformed nuclei also are suggested to be capable of vibrations about an equilibrium deformed shape.  
  The emergence of predominantly prolate spheroidal shape moments suggested two vibrational modes in deformed nuclei:  ``gamma'' vibrations ($Y_{22} + Y_{2-2}$ multipole mode) and ``beta'' vibrations ($Y_{20}$ multipole mode).  
  
  There is a voluminous literature that discusses one-phonon $\beta$-vibrational and $\gamma$-vibrational states in deformed nuclei; and the lowest-lying excited $K^\pi = 0^+$ and $2^+$ states, respectively, are generally identified with these modes.  
  However, vibrations in quantum systems should exhibit multi-phonon eigenstates.  
  Evidence for multiple (two) phonon excitations in deformed nuclei is sparse and has been difficult to obtain.  
  
  The best examples for multiple phonon excitations are limited to evidence for two-phonon $\gamma$ vibrations, but controversy over this structural interpretation persists (see, e.g., \cite{Borner1991, Burke1994, Wu1994, Oshima1995, Fahlander1996, Garrett1997, Hartlein1998}).
  That the double gamma vibration is hard to identify is connected to the high level density in well-deformed nuclei and the low spin of the states of interest.
  In contrast to the gamma vibration, whose excitation energy generally decreases with mass, the beta vibration is expected to have the lowest energy in lighter nuclei, which have a lower level density.
  There is no unequivocal evidence for two-phonon $\beta$ vibrations.  

  The nucleus $\sm$ is particularly well-suited as a case study for the existence of multi-phonon $\beta$ and $\gamma$ vibrations in a deformed nucleus.  
  It has one of the lowest-energy candidate $\beta$ vibrations in any deformed nucleus (and a fairly low-energy candidate $\gamma$ vibration), such that 2- and even 3-phonon excitations should be below the pairing gap, if they exist.
  Indeed, recently it has been suggested \cite{Garrett2001} that $\sm$ and its neighboring isotone, $\gd$, are the best candidates for establishing the $\beta$-vibrational mode in deformed nuclei.  
  It has long been regarded as a ``soft'' nucleus \cite{Mackintosh1977}.

  To address the expectation that these simple multi-quantum excitation modes should exist in $\sm$, we have carried out a very-high-statistics study using multiple-step Coulomb excitation (multi-Coulex).  
  This study was made using the Gammasphere array \cite{Lee1996}  of Compton-suppressed Ge detectors in conjunction with the CHICO charged-particle detector array \cite{Simon2000}.  
  The experiment used a beam of $\sm$ ($E=652$~MeV, an energy insufficient to surmount the Coulomb barrier) incident on a thin $\pb$ target (400 $\mu$g/cm$^2$, 99.86$\%$ enrichment) at the Lawrence Berkeley National Laboratory's 88-Inch Cyclotron.  
  Signals from two ions detected by CHICO in coincidence with at least one ``clean'' $\gamma$ ray signal in Gammasphere (i.e., a p-p-$\gamma$ coincidence) triggered an event.  
  The CHICO array provided kinematic characterization of scattered ions and recoiling target nuclei for Doppler corrections to the $\gamma$ rays emitted from the Coulomb-excited beam nuclei.  
  High angular resolution was provided both by CHICO, which has 1$^{\circ}$ angular resolution, and by Gammasphere, which was operated with 104 Ge detectors.  
  A total running time of 62 hours provided 7 $\times$ 10$^8$ p-p-$\gamma$, 8 $\times$ 10$^7$ p-p-$\gamma$-$\gamma$, and 1 $\times$ 10$^7$ p-p-$\gamma$-$\gamma$-$\gamma$ events.

\begin{figure}[htbp]
\resizebox{8.6 cm}{!}{\includegraphics*{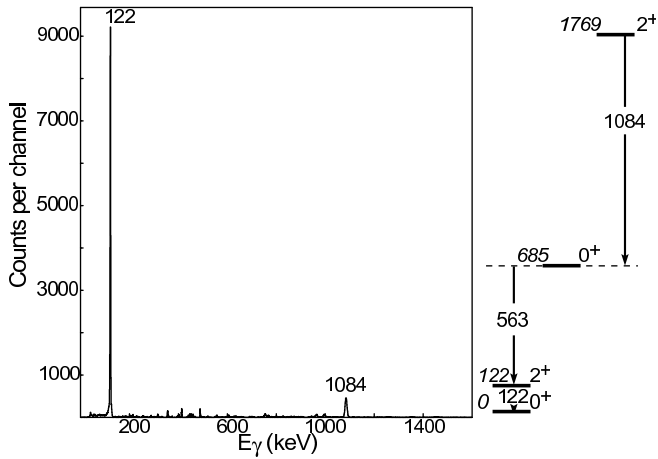}}
\caption{\label{563 gate} Spectrum of $\gamma$ rays in coincidence with the $0^+_2$ (685) $\rightarrow 2^+_1$ (122), 563 keV $\gamma$-ray transition in $\sm$.  
  The intense 122~keV line is the $2^+_1$ (122) $\rightarrow 0^+_1$ (0) ground state band transition below the gate transition.
  The 1084~keV $\gamma$ ray from the $2^+$ state at 1769~keV is the only other prominent feature in the spectrum.}
\end{figure}

  If a two-phonon excitation exists at low energy in $\sm$, it would be evident through $\gamma$-ray transitions coincident with $\gamma$ rays de-exciting the $K^\pi = 0^+$ rotational band built upon the [$J^\pi_i$ ($E_x$ keV)] $0^+_2$ (685) state, the lowest excited (non-rotational) structure in this nucleus. 
  Figure \ref{563 gate} shows a p-p-$\gamma$-$\gamma$ coincidence gate on the $0^+_2$ (685) $\rightarrow 2^+_1$ (122), 563~keV $\gamma$-ray transition.
  The only strong lines in this spectrum are the known $2^+$ (122) $\rightarrow 0^+_1$ (0), 122~keV and $2^+$ (1769) $\rightarrow 0^+_2$ (685), 1084~keV transitions in $\sm$ \cite{Artna-Cohen1996}.
  The simplicity of the gated spectrum indicates that there is little, if any, fragmentation of collective strength connected to the $0^+_2$ (685) state and that this strength is concentrated in an excitation which has the $2^+$ (1769) state as a band member.

\begin{figure}[htbp]
\resizebox{8.6 cm}{!}{\includegraphics*{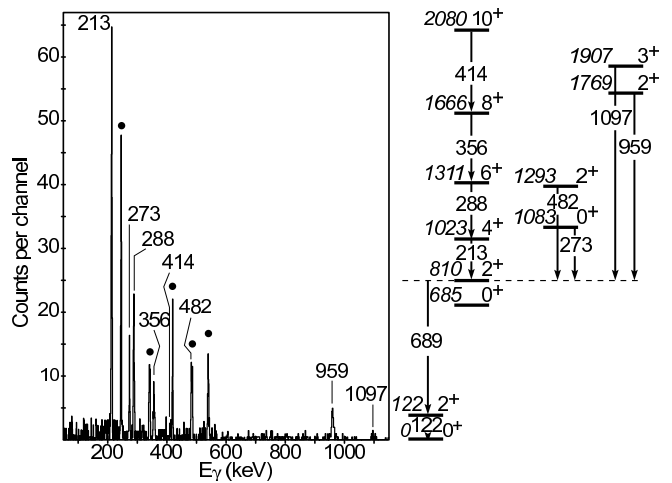}}
\caption{\label{689 gate} Double-coincidence gate indicating $\gamma$ rays in coincidence with the $2^+_2$ (810) $\rightarrow 2^+_1$ (122) $\rightarrow 0^+_1$ (0), 689 and 122 keV cascade $\gamma$-ray transitions in $\sm$.  
Lines indicated by a dot above the $\gamma$ ray are due to an overlap with the $13^-_1$ (2833) $\rightarrow 12^+_1$ (2149) $\rightarrow 2^+_1$ (122) $\rightarrow 0^+_1$ (0), 685 and 122~keV double-coincidence gate.}
\end{figure}

  Figure \ref{689 gate} shows a p-p-$\gamma$-$\gamma$-$\gamma$ coincidence event gated using  the $2^+_2$ (810) $\rightarrow 2^+_1$ (122) 689~keV and (122) $\rightarrow 0^+_1$ (0) 122~keV $\gamma$-ray transitions.
  In this spectrum, the $\gamma$ rays at 213, 288, 356, and 414~keV correspond to in-band transitions higher up in the rotational band built upon the  $0^+_2$ (685) state \cite{Artna-Cohen1996};
    the lines marked with solid dots are transitions in the ground-state band that are due to 685-122 coincidences, where the 685~keV $\gamma$ ray is a known $13^-_1$ (2833) $\rightarrow 12^+_1$ (2149) transition \cite{Artna-Cohen1996};
  and the relatively weaker $\gamma$ rays at 273 and 482~keV arise from the pairing isomeric band built on the $0^+_3$ 1083~keV state \cite{Kulp2005a, Kulp2007a}.
  Above 550~keV there are only two transitions, both of which directly feed the $2^+_2$ (810) state:  one line at 959~keV and another at 1097~keV (discussed later).
  The 959~keV $\gamma$ ray also de-excites the $2^+$ (1769) state \cite{Artna-Cohen1996, Kulp2007a}, and as this level is higher in energy than many other known excited states in $\sm$, the question arises as to what kind of excitation is involved.
  In particular, is it the $2^+$ member of a $K^\pi = 0^+, 1^+$, or $2^+$ band?

  To thoroughly explore excitations in $\sm$ in this energy range, we complemented the multi-Coulex study with an $(n,n'\gamma)$ study of $\sm$ at the University of Kentucky with monoenergetic neutrons.
  The data obtained included excitation functions ($E_n$ in 0.1 MeV increments from 1.2 to 3.0 MeV), $\gamma$-ray angular distributions, Doppler-shifted $\gamma$-ray energy profiles, and $\gamma-\gamma$ coincidences (details of similar analyses may be found in \cite{Belgya1996}).
  These data provide comprehensive information on low-spin states, including spins, decay branches, and lifetimes to $\ge 2.1$ MeV excitation.
  Here we focus on low-spin states of positive parity, and a first result is that no $1^+$ states are observed below 2 MeV, eliminating the possibility that the $2^+$ (1769) level is a $K^\pi = 1^+$ state.
  It is then a question of whether the $2^+$ (1769) level is a member of a $K^\pi = 0^+$ band or the band-head of an excited $K^\pi = 2^+$ band.
  
\begin{figure}
\resizebox{8.6 cm}{!}{\includegraphics*{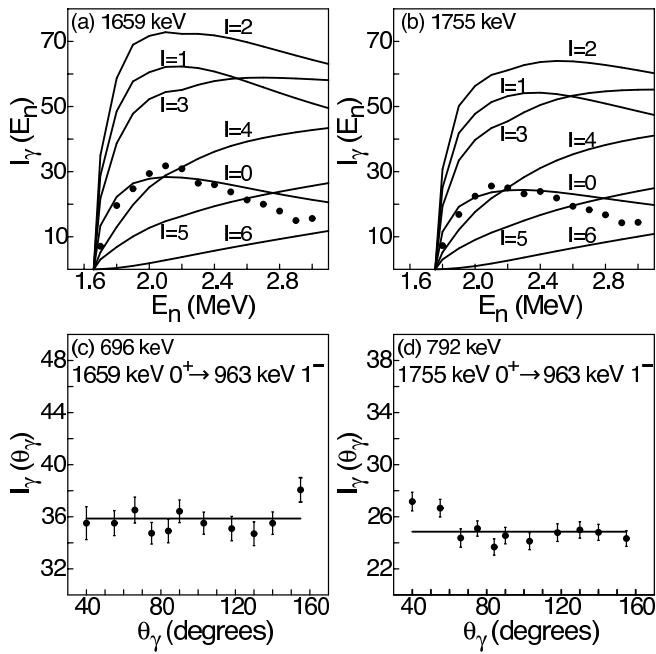}}   
\caption{\label{excit} Excitation functions and angular distributions for $\gamma$ rays that depopulate the $0^+_4$ 1659 and $0^+_5$ 1755 keV levels, respectively.}
\end{figure}

\begin{figure}
\resizebox{8.6 cm}{!}{\includegraphics*{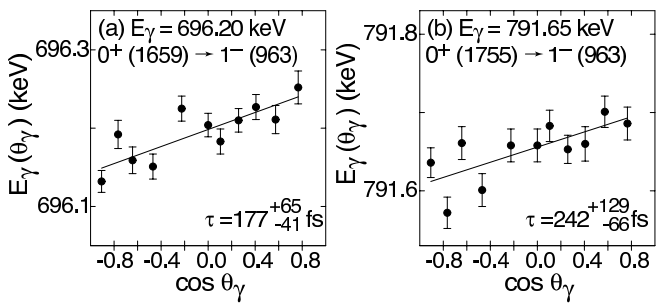}}
\caption{\label{lifetimes}  Doppler shift plots of the a)  696 and b)  792 keV $\gamma$ rays following the ($n,n'\gamma$) reaction on $\sm$.  Lifetimes of the $0^+_4$ 1659 and $0^+_5$ 1755 keV levels were extracted from these data.}
\end{figure}

  If the $2^+$ (1769) state has $K^\pi = 0^+$, then there should be a $0^+$ state 100-200~keV below it.
  The previously known \cite{Artna-Cohen1996} excited $0^+$ states in $\sm$ are at  684.7, 1082.9 (a pairing isomer \cite{Kulp2005a}), and 1659.5 keV.
  We confirm that these states have $0^+$ spin-parity and establish a new $0^+$ state at 1755.0~keV.
  The 1659.5 and 1755.0~keV $0^+$ states lie below the $2^+$ (1769) state and are therefore of particular interest.
  Excitation function data for these states are shown in Fig.~\ref{excit}.
  These data can be compared with the plotted curves of theoretical direct population of the levels as a function of neutron energy and level spin, calculated with the CINDY computer code \cite{CINDY1973} using input optical model parameters from the RIPL-2 database \cite{RIPL2}.
  The best agreement between plotted data and theoretical curves indicates the 1659.5 and 1755.0~keV levels are spin-0 states, and this is confirmed by the isotropic $\gamma$-ray angular distributions in the lower part of Fig.~\ref{excit}.
  From Doppler-shifted energies of decaying $\gamma$ rays (shown in Fig.~\ref{lifetimes}), lifetimes of $177^{+65}_{-41}$ and $242^{+129}_{-66}$~fs are determined, respectively, for the 1659.5 and 1755.0~keV states.
  We deduce $B(E2; 0^+, 1659 \rightarrow 2^+_2, 810) = 5.1^{+1.0}_{-1.5}$ W.u.\ and $B(E2; 0^+, 1755 \rightarrow 2^+_2, 810) < 5$ W.u.\ using decay paths established in the present work and in \cite{Artna-Cohen1996}.

\begin{figure}
\resizebox{8.6 cm}{!}{\includegraphics*{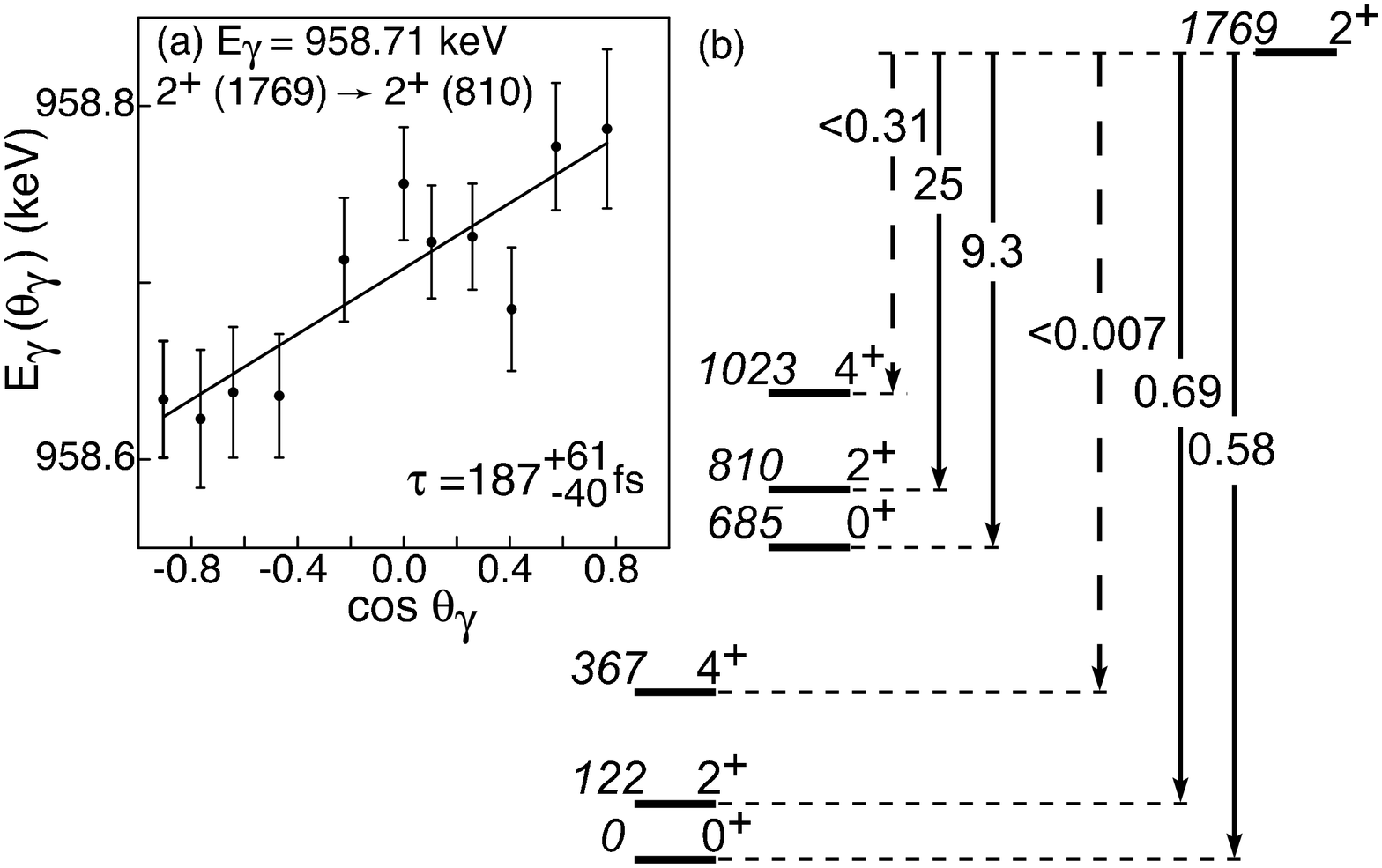}}
\caption{\label{1769 level}  a)  The lifetime of the 1769~keV $2^+$ state was extracted from the Doppler-shifted energy as a function of angle of the 959 keV $\gamma$ ray following the ($n,n'\gamma$) reaction on $\sm$.  
  b)  Deduced $B(E2)$ values in W.~u.\ for transitions from the 1769~keV state to the lowest two $K^\pi = 0^+$ bands.}
\end{figure}

  We now consider the $E2$ decay strength exhibited by the $2^+$ (1769) state.
  Figure \ref{1769 level}a shows a plot of the Doppler-shifted energy for the 959~keV $\gamma$ ray, from which we deduce a lifetime for the 1769~keV level of $187^{+61}_{-40}$ fs.
  Assuming pure $E2$ character for transitions between the 1769~keV level and states in the lowest two $K^\pi = 0^+$ bands, the $B(E2)$ values deduced for these $\gamma$ rays from the measured lifetimes and $\gamma$-ray intensities are shown in Fig.~\ref{1769 level}b.
  Evidently, the 1769~keV level has a strong collective connection with the $K^\pi = 0^+$ excited band built on the $0^+$ (685) state, as previously illustrated by the strong 1084~keV (9 W.u.) and 959~keV (25 W.u.) $\gamma$ rays in Figs.~\ref{563 gate} and \ref{689 gate}, respectively.
  Upper limits of $B(E2; 1769 \rightarrow 367) <0.007$ W.u.\ and $B(E2; 1769 \rightarrow 1023) < 0.31$ W.u.\ are determined using previously unpublished data from our recent $\eu^{g}$ study \cite{Kulp2007a}.
    
    Comparing the $B(E2)$ data in Fig.~\ref{1769 level} with expected relative $B(E2)$ values from the Alaga rules, we can determine $K^\pi$ for the 1769~keV state.
    The Alaga rules for $\Delta K=0$ transitions from a $J^\pi=2^+$ state to states of $J^\pi=0^+/2^+/4^+$  indicate that $B(E2)$ ratios of 70/100/180 are expected.
    In contrast, $B(E2)$ ratios of 70/100/5 would be expected for $\Delta K=2$ transitions.
    The experimental data clearly show best agreement with the $\Delta K = 2$ expectations, and therefore the $2^+$ (1769) state is the band-head of a second-excited $K^\pi = 2^+$ band.

  A $K^\pi = 2^+$ assignment for the $2^+$ (1769) state is further supported by the observation of the 1097~keV $\gamma$ ray in Fig.~\ref{689 gate}.
  This line de-excites a level at 1907~keV, which is established in both the multi-Coulex and the $(n,n'\gamma)$ studies and has a probable spin-parity of $3^+$.
  
\begin{figure}[htbp]
\resizebox{8.6 cm}{!}{\includegraphics*{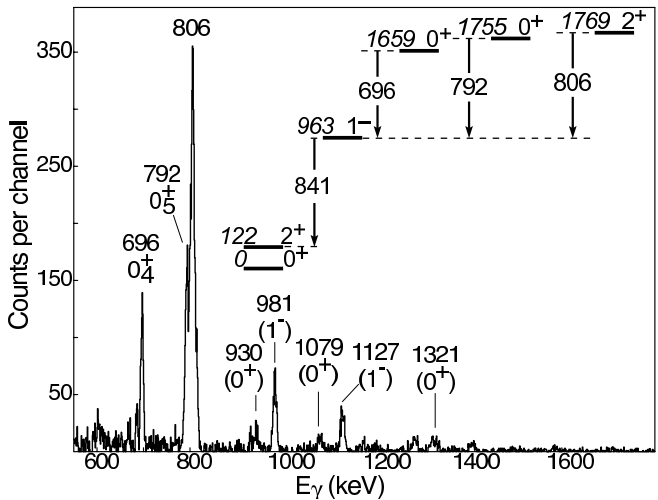}}
\caption{\label{841 gate} Coincidence gate on the $1^-_1$ (963) $\rightarrow 2^+_1$ (122), 841 keV $\gamma$ ray in $\sm$.  Details of the observed transitions are discussed in the text.}
\end{figure}

  A broader view of low-spin collective states in $\sm$ is provided in Fig.~\ref{841 gate}.
  This figure shows $\gamma$ rays from the p-p-$\gamma$-$\gamma$ data in the multi-Coulex study observed in coincidence with the $1^-$ (963) $\rightarrow 2^+$ (122) 841~keV transition.
  The very strong line at 806~keV de-excites the $2^+$ (1769) level \cite{Artna-Cohen1996,Kulp2007a}.
  The 696 and 792~keV $\gamma$ rays de-excite the $0^+$ states, discussed above, at 1659 \cite{Artna-Cohen1996} and 1755~keV.
  Other lines are weak and are identified with the de-excitation of $J=0, 1$ states known from $\Pm$ $\beta^-$ decay \cite{Artna-Cohen1996} or established in our $(n,n'\gamma)$ study.
  The pattern of strength is qualitatively a reflection of the $E2$ strengths associated with the $0^+$ (1659) \cite{Artna-Cohen1996}, $0^+$ (1755), and $2^+$ (1769) states, as established in this work.
  No other low-spin ($J\le2$) collective states, besides these three, are indicated in $\sm$ below $\sim$2750~keV. 

  The absence of an excited $K^\pi = 0^+$ structure with strong $E2$ transitions to the $K^\pi = 0^+$ structure based on the 685~keV state is very surprising. 
  The conventional view of the $0^+$ (685) state in $\sm$ has been that it is an intrinsic quadrupole excitation of the ground state, i.e., a $\beta$ vibration.
  Alternative theoretical descriptions by the interacting boson model (IBM) \cite{Zamfir1999a,Klug2000}, pairing-plus-quadrupole model (PPQ) \cite{Kumar1974}, and X(5) model \cite{Iachello2001, Casten2001a} also fail to explain our data:  These models predict second excited $0^+$ states at 1279, 1473, and 1729~keV that have strongly collective transitions to the $2^+$ (810) state with $B(E2)$ values of 103, 82, and 122 W.u., respectively.
  The experimental data presented here show that the existence of such intrinsic excitations in $\sm$ are excluded.
  
  In summary, we have carried out multiple-step Coulomb excitation and inelastic neutron scattering studies of $\sm$ to explore the possibility of multiple-phonon intrinsic collective excitations in this nucleus.
  Four excited $0^+$ states below $\sim$1.8 MeV have been identified, but no evidence for two-phonon $K^\pi = 0^+$ quadrupole vibrations was found.
  A second excited $K^\pi=2^+$ band based on the $2^+$ (1769) state was identified, and the collective nature of transitions from the 1769~keV level to members of the band built on the $0^+_2$ (685) state was determined through a measurement of the lifetime of this level.
     
  The results of this work show that a new interpretation for the $0^+$ (685) state is called for.
  Further, the collective connection between the $K^\pi = 0^+$ band built on the $0^+_2$ (685) state and the $K^\pi = 2^+$ band built on the $2^+$ (1769) state appears to be unprecedented in nuclei in this mass region.
    
\begin{acknowledgments}
  We wish to thank colleagues at the LBNL 88-Inch Cyclotron and the University of Kentucky monoenergetic neutron facility for their assistance in these experiments.  
  This work was supported in part by DOE grants/contracts
DE-FG02-96ER40958 (Ga Tech), DE-AC03-76SF00098 (LBNL),  and by NSF awards PHY-0244847 (Rochester) and PHY-0652415 (Kentucky).

\end{acknowledgments}


\end{document}